\documentclass[12pt]{article}

\usepackage{slashed}

\newcommand{\bpsi}{\overline{\psi}}

\begin{document}

\title{ Higher order fermion effective interactions
 in a 
bosonization approach
}

\author{Fabio L. Braghin
\\
Instituto de F\'\i sica, Fed. Univ. of  Goias,
\\
 P.B.131, Campus II, 
74.001-970, Goi\^ania, GO, Brazil
}

\date{Received: date / Accepted: date}

\maketitle

\begin{abstract}
 Three different  fermion effective potentials
 given by series of bilinears, 
 $\sum_j^N  (\bar{\psi}_a \psi_a)^{2^j}$ , 
$\sum_j^N   (\bar{\psi}_a \psi_a)^{j}$ 
and also  
 $\sum_j^N   (\bar{\psi}_a \gamma_\mu \psi_a)^{2j}$ where  $a=1,...N_r$
and integer $j$
are investigated by introducing  sets of auxiliary fields.
A mininal procedure is adopted  to  deal  with the auxiliary fields
and an effective bosonized model 
 in each case is found by assuming  
weak  field  fluctuations, i.e.  weak enough when
  compared to  (normalized) coupling constants.
Different fermion condensates are considered for the ground state 
 in the first  two  series   analysed
and  the  factorization of all higher order condensates into  the lowest order 
one  is found in most cases,   i.e. in general 
 $<(\overline{\psi}_a \psi_a)^n> \propto <\bpsi_a \psi_a>^n$.
For the case of the third series built with vector-type bilinears
no condensation is assumed to occur.
The corresponding (weak) scalar fields
 effective models for the three cases
are expanded in  polynomial interactions.
The resulting low energy 
effective boson  model may a exhibit new approximate  symmetry
depending on the terms present in the original series-model and on the 
values of the coupling constants.
\end{abstract}

\section{Introduction}

Higher order polynomial interactions 
usually appear
in effective field theories
including in cases in which
  non-polynomial interactions  might be expanded into 
series of polynomial interactions
\cite{qft,osipov-hiller1,osipov-hiller2,osipov-hiller-blin,simmons1,simmons2,chalmers,nitta-susy,darkmatter,cahill,6th-order-cm,braaten-hammer,bedaque-kolck,weinberg79,ChPT1,ChPT2,kleinert,e-w-EFT}.
Although  higher order interactions
are usually  irrelevant from a renormalization group analysis,
 one can ask  whether and how they can   contribute separatedly
 to the ground state and 
dynamics of the system.
Further 
 technical difficulties arise
 requiring  reliable approximative methods to treat them.
Higher order effective interactions  contribute, 
in particular, in the framework of  the 
Effective Field Theories (EFT) 
that has been shown to be suitable for the  problem of  (few) N-body states 
\cite{braaten-hammer,bedaque-kolck}.
The present work is to be seen as a non perturbative approach to these problems
being that  it decomposes the higher order effective fermion
 potential into N-body states.
In this perspective, the factorization problem mentioned below 
may have consequences  in   few body
systems  since
some 3 and 4 body states are usually described only  in terms of the 2 and 
3 body scales and parameters 
\cite{platter-etal}.
 Quantum Chromodynamics and the Electroweak theory are emblematic 
 examples of   theories whose effective models are of high interest for 
particular sectors of the corresponding phase diagrams.
For instance,
one of the most well known  examples for higher order effective
interactions comes from the low energy Quantum Chromodynamics (QCD)
in  which   series of higher order  couplings 
 appear  \cite{weinberg79,ChPT1,ChPT2,kleinert}.
Although the present work does not address 
QCD and their effective models,
multifermion states and condensates 
($<(\bpsi_a \psi_a)^n>$, for $n=1,2,...$) usually considered
in QCD
can also be considered for different fermion models when a 
spontaneous symmetry breaking takes place.
In QCD, due to  difficulties associated with calculating higher order 
multiparticle states and condensates, it was proposed
a factorization  hypothesis 
of  the higher order condensates  into the lowest order one
 (i.e.  $<(\bar{q} q)^{m}> \sim (<\bar{q}q>)^{m}$)  \cite{factorization,zong-etal}.
However one finds 
it is not  a very good approximation
since 
$<(\bar{q} q)^{m}> \neq (<\bar{q}q>)^{m}$ 
 \cite{nonfactorization,braghin-navarra}.
This issue of higher order condensates, and their (non)factorization,
 might be  however a general problem for  fermion 
 quantum field theories undergoing certain spontaneous symmetry breakings.
One of the aims of the present work is  to investigate the very low energy regime of general fermion
effective models
with higher order interactions
whenever  composite  fermion states and condensates are formed.
How and under which circumstances 
condensation occurs will not be discussed.
Although these models are easily made
invariant under U(1), U(1)$^{N_r}$ and  U($N_r$),
being that all bilinears $(\bpsi_a \psi_a)$ have an implicit summation over $a$,
 the role of these symmetries will not be specifically investigated here.
Further
motivations for the present work can be found in 
systems where   n-body states (mainly $n=2$ and $3$ or $4$)
arise with some approximate degeneracy among
composite states with different number of particles,
in particular 
in cold atoms \cite{cold-atoms1,cold-atoms2} where (non relativistic) 
2, 3 and 4 boson or fermion states are formed
and in (light scalar) meson spectroscopy \cite{scalars1,scalars2} where
(relativistic) light mesons with quark-antiquark and  tetraquark 
structures have similar masses.

The   non-perturbative  auxiliary field method
is 
known to provide good results for small fluctuations and weak coupling constant 
 being directly   extended to incorporate 
loop perturbative corrections \cite{kashiwa}.
In spite of the difficulties associated to a more complete and exact account 
of the nonlinearities \cite{kleinert2011}
it has been widely and successfully  applied to different models such as 
Gross Neveu, Nambu-Jona-Lasinio, among other models
 \cite{GN,klevansky,berdnorz,lp4,nitta-susy,condmat}.
This  method can be implemented by means
 of shifts of the auxiliary fields to produce interactions 
that cancel out the original interactions of the model yielding 
a linearization of the original  Lagrangian.
A similar procedure given in Ref. \cite{alkofer-reinhardt}
also  produce an effective boson model for  the original fermion model.
In this work, and in Ref. \cite{braghin-navarra},
the auxiliary field method 
is considered   to incorporate   the corresponding 
higher order fermion composite states and condensates ($<(\bpsi_a \psi_a)^n>$)
by means of suitable shifts of extra (higher order) auxiliary fields. 
With the present approach however it is possible to envisage if
these higher order condensates factorize or not into the lowest order one.

In this work,  three different  effective fermion models are investigated
with interactions given by simple  series of  bilinears 
$\sum_n (\bpsi_a \psi_a)^{2^n}$, 
 $\sum_n (\bpsi_a \psi_a)^n$ and $\sum_j   (\bar{\psi}_a \gamma_\mu \psi_a)^{2j}$,
for $n \leq N$ or $j \leq N$, and 
where $a=1...N_r$ is an internal quantum number.
This work only deals with finite $N$ effective interactions, in particular with 
a maximum power $N_{max} = 4 \times N_r$ in four spacetime dimensions,
since for $N >  N_{max}$ there is no non trivial fermion effective interaction
due to the Grassmann algebra. 
 These series can  be  considered to be simply  toy or test  models or effective models for 
more fundamental theories.
Strictly speaking, a
 renormalization group (RG) evolution for the more fundamental
theory  generates all the 
missing bilinears in the first 
series: 
$(\bar{\psi} \psi)^m$ with powers $m< 2^n$.
At some energy scale  it
 is possible that some of the generated terms are relatively small when compared to 
the
others, i.e. some effective coupling constants (those of the terms $g_{2^n}$)
are relatively larger than the others,
by 
considering they 
have different dimensions.
Alternatively, one may simply consider the first series 
(with the terms $g_{2^n}$) as a mathematical limit of the more general series
and try to understand the role of the terms of the more general series 
that are not included.
In these cases
(models either realistic due to the original fundamental  interactions
or as a limiting mathematical case of the general model), 
it is legitimate to analyse 
 a resulting effective model with such 
series of bilinears.

Therefore,  as discussed above, the aims of this work are the following.
Firstly,  a dynamical framework will be considered
to  investigate if the higher order condensates 
 factorize or not 
(whenever they can be formed)
 into the lowest order condensate, 
i.e. if they behave as
$<(\bpsi_a \psi_a)^n> \propto <\bpsi_a \psi_a>^n$ or not, up to $n=N$
and for $a = 1...N_r$.
This is done by articulating further the 
 non perturbative  auxiliary field method to treat higher order interactions self consistently,
hopefully it can  bring some insight/information 
on the  role of such effective interactions  in the low energy dynamics.
This is done by building  N-body (composite boson) states 
(built from  N-fermion states)
taking into account (the corresponding) 
N-body effective interactions in a dynamical framework.
Finally, we wish  to extract information about the contribution of higher order fermion interactions
for the low energy dynamics for models in which there is no
invariance under chiral symmetry ($U(N)\times U(N)$ or $SU(N)\times SU(N)$), 
which is an important symmetry in   low energy Chromodynamics.
In this way it is possible to provide hints or results about resulting properties 
that are not due to  chiral symmetry.
To provide the cancelations of the interactions, 
   shifts are performed in the normalized 
Gaussian integrals within  the standard auxiliary field method. 
A minimal procedure is adopted 
which requires  the minimum  number of auxiliary fields
and 
the minimum number of 
shifts, preventing the appearance of  ambiguities.
These fields are assumed to be  weak when compared to the mean fields.
and few ways of extending the validity of the 
results are  shown,  i.e. for non necessarily weak  fields.
 yielding essentially unchanged  results.
The second aim, is to  
 expand  the resulting effective model for the auxiliary  fields
 to show the structure of the resulting
  polynomial  effective  model. 
Comparison of the results from the first (or third)  series of interactions
with the more general second series 
 will show   an  analytical example of an extra  symmetry for the 
resulting 
effective boson model.
The article is organized as follows. 
In the next section the method is presented for a series of interactions
of the type $\sum_n (\bpsi_a \psi_a)^{2^n}$,   the  ground state gap 
equations   of  auxiliary fields
are shown and the (secondary level) polynomial effective model is derived.
A way to lift the weak field approximation is presented in Sec \ref{sec:no-weak}.
In Sec. \ref{sec:geral} a more general series, of the type  $\sum_n (\bpsi_a \psi_a)^n$,  is investigated
within the  same procedure of Section \ref{sec:esp}.
In Sec. \ref{sec:vector} 
a model with interactions  typical from a momentum independent local limit of the effective potential
obtained by
the exchange of a vector field ($\sum_j^N   (\bar{\psi}_a \gamma_\mu \psi_a)^{2j}$)
is considered without the formation of condensates (mean fields).
In the final section there is a summary of the results.

\section{ Series of interactions $\sum_n (\bpsi_a \psi_a)^{2^n}$}
\label{sec:esp}

Consider the generating functional of an effective model for fermions
$Z =  \int {\cal D} [\bpsi_a, \psi_a]
e^{  \; i \int_x {\cal L} [\psi_a, \bpsi_a]} $,
where $\int_x$ denotes space-time integration in d-dimensions, with the Lagrangian density given  by:
\begin{eqnarray} \label{series-I}
{\cal L} &=& 
\bpsi_a (x) \left( i \slashed{\partial} - m_a \right) \psi_a (x)
+ \sum_n^{N}  g_{2^{n}} \left( \bpsi_a \psi_a \right)^{2^n } , 
\end{eqnarray}
where $g_{2^n}$ are the effective coupling constants with 
dimension: $[g_{2^n}] = M^{d- (d-1)2^n}$, where $M$ has dimension of mass,
$m_a$ are the masses for each of the fermion species
and the index 
$a=1...N_r$ stands for the fermion components
being that in
 each bilinear has an implicit summation over $a$ 
and the mass term is therefore diagonal.
The fermion interactions will be eliminated in favor of  a set of 
scalar  auxiliary fields 
which might  give rise to  
the scalar structures  of the type $[(\bpsi_a \psi_a)^n]$.

These auxiliary  fields  (a.f.) are introduced by means of the  following unity
 integrals  multiplying the generating functional:
\begin{eqnarray} \label{gaussian}
N'  \int \; {\cal D} [\varphi_n]
 e^{- i \int_x \frac{1}{2} \sum_n^{N} \frac{1}{d_n} \varphi^2_n (x)} 
 = 1 ,
\end{eqnarray}
where $N'$ is a normalization constant and the parameters 
$d_n$ are left free for the sake of generality.
Alternatively a  rescaled set of auxiliary fields could have been chosen:
$
\frac{1}{d_n} \varphi_n^2  = \tilde{\varphi}_n^2$,for 
$\tilde{\varphi}_n = \frac{1}{\sqrt{d_n}} \varphi_n
$
where the parameters could have  been chosen to be simply $d_n = 1$.
The necessary shifts of the a.f. needed to cancel out all the interactions 
can be made minimal shifts, i.e.,  the simplest shifts for the minimum number of
auxiliar fields which do not introduce Lagrangian terms  that  were not presented
in the original model. 
For the model of expression (\ref{series-I}) the shifts  are given by:
\begin{eqnarray} \label{shifts-4n-1}
\varphi_n^2 \to (\varphi_n - \beta_n (\bpsi_a \psi_a)^{2^{n-1}} 
)^2,
\end{eqnarray}
where $\beta_n$ are dimensionful parameters that are determined 
by imposing the corresponding cancelations of all polynomial  interactions.

To obtain  a finite number of equations,
let us consider the series ends at $n=N=5$, 
 being
easily generalized.
The conditions for the cancelations of the polynomial interactions
are the following:
\begin{eqnarray}
g_{32} &=& \frac{\beta^{2}_{5}}{2 d_{5}} ,
\nonumber
\\
g_{16} &=&  - \frac{2 \beta_5}{2 d_{5}} \varphi_5 + 
\frac{\beta^{2}_{4}}{2 d_{4}} ,
\nonumber
\\
g_8 &=& - \frac{2 \beta_4}{2 d_{4}} \varphi_4
+ \frac{\beta_3^2}{2 d_3} ,
\nonumber
\\
g_4 &=& - \frac{2 \beta_3}{2 d_3} \varphi_3
+ \frac{\beta_2^2}{2 d_2} ,
\nonumber
\\
g_2 &=&  - \frac{2 \beta_2}{2 d_2} \varphi_2
+ \frac{\beta_1^2}{2 d_1} .
\end{eqnarray}
For an arbitrary $n$, these conditions can be written in the following form:
\begin{eqnarray} \label{shift-4n-geral}
g_{2^n} &=&  - \frac{\beta_{n+1} }{ d_{n+1}} \varphi_{n+1}
+
\frac{\beta_n^2}{2 d_n} 
,
 \;\;\;\;  \mbox{for n $<$ N} ,
\nonumber
\\
g_{2^n} &=& 
\frac{\beta_n^2}{2 d_n} , \;\;\;\;\;\;\;\;\;\;\;\;\;\;\;\;\;\;\;\;\;\;\;\;\;\;\;\;\;
\mbox{for n $=$  N}.
\end{eqnarray}
If
 the parameters $\beta_n$ are then considered to be functions of different a.f.
the above conditions 
 one  must garantee that the shifts of the a.f. still have
unity Jacobian. 
In fact all these shifts yield  $\beta_n = \beta_n [\varphi_{n+1}]$ and
 these still yield  unity Jacobian.
In fact, different shifts that could cancel out the 
fermion interactions would  introduce
other non linearities and the need of extra a.f. or 
non unity Jacobians.
These parameters will  assume   numerical values 
 (for fixed  values for the coupling constants
$g_n$) when solving the gap equations,
and then $\beta_n \to \beta_n[\varphi_{n+1}^{(0)}]$.
It yields the following relations:
\begin{eqnarray} \label{betas}
\beta_n &=& \sqrt{2 d_n \left( g_{2^n} + \frac{\beta_{n+1} }{d_{n+1} } \varphi_{n+1}
\right) }     \;\;\;\;   \mbox{for n $<$ N},
\nonumber
\\
\beta_n &=& \sqrt{2 d_n g_{2^n} }    \;\;\;\;\;\;\;\;\;\;\;\;\;\;\;\;\;\;\;\;\;\;\;\;\;\;\;\;\;\;
 \mbox{for n $=$ N}.
\end{eqnarray}
In this case
there is no ambiguity in the definitions of the parameters
$\beta_n$ as functions
of the a.f. $\beta_i[\varphi_{j+1}]$ written above.
This minimal procedure   is  valid when all the fermion coupling constants, 
except $g_{2^N}$, are quite strong 
and  
(1) a subset of  $\varphi_{N-1}$ fields only assume positive values
or 
(2) these auxiliary fields are weak with respect to 
the mean field which are weaker than the (normalized) fermion coupling constants.
This means that higher order auxiliary fields, which are introduced to 
cancel out progressively more irrelevant fermion interactions,
are  progressively weaker,
 i.e.  
$|\varphi_m \beta_m | <  g_{2^{m-1}}$ (positive coupling constants)
where $\varphi_m$ is the mean field plus the fluctuation.
Some ways to lift  these conditions of weak field regime are  provided 
in the next section.

The resulting effective  action is given by:
\begin{eqnarray} \label{Seff-n}
S_{eff} &=&  \int d^4 x \;  
\left[
\bpsi_a\left( i \slashed{\partial} - m_a + \frac{\beta_1}{d_1} \varphi_1 
\right) \psi_a 
- \sum_{n=1}^N \frac{1}{2 d_n} \varphi^2_n \right].
\end{eqnarray}
The saddle point equations for these auxiliary fields 
 provide   relations between the 
ground state average of the auxiliary fields $\varphi_n$ and 
the progressively large powers of bilinears
$< ( \bpsi_a \psi_a )^n>$.
To show  these relations,  consider that 
$\frac{\delta \beta_1}{\delta  \varphi_n}  =
\left( \prod_2^{n-1} \frac{\varphi_i}{\beta_i}
\right) \frac{d_1}{d_n} \frac{\beta_n}{\beta_1}$.
 It yields:
\begin{eqnarray} \label{def-cond-I}
\frac{ < \varphi_n > }{\beta_n}  \equiv \frac{\varphi_n^{(0)} }{\beta_n} 
\equiv < (\bpsi_a \psi_a)^n >.
\end{eqnarray}
The auxiliary fields analysed in this work are all scalars, i.e. they
encapsulate  the combinations of the fermion bilinears for each species,
namely:
$\varphi_n \sim (\bpsi_a \psi_a)^n$
for implicit summation over $a$.
Therefore these auxiliary  fields   represent 
a sum of (physical) states.

By integrating out fermions, 
the following effective action is obtained:
\begin{eqnarray} \label{Seff-n}
S_{eff} = -  i Tr \log \left( i \slashed{\partial} - m_a + \frac{\beta_1}{d_1} \varphi_1 
\right) - \sum_{n=1}^N \int_x \frac{\varphi^2_n}{2 d_n},
\end{eqnarray}
where $Tr$  is the traces taken over all the internal
 indices and integration over space-time.
According to expression (\ref{betas}), there is a   dependence of
$\beta_1$ on all the fields $\varphi_n$ through the parametric dependence of 
$\beta_1$ on $\beta_n$ (for $n\neq 1$), i.e.:
\begin{eqnarray}
 \beta_1  = \beta_1 [ \varphi_2, \beta_2 ]
\to \beta_1 [\varphi_2 [ \varphi_3 [... [\varphi_N]]]].
\end{eqnarray}
Therefore $\beta_1$ in the effective mass encodes  the non linearities of the model.

The resulting mean field (homogeneous) GAP equations are the following:
\begin{eqnarray} \label{gap-n-1}
\frac{\varphi_1}{d_1} &= -  i  \frac{\beta_1}{d_1} 
Tr \frac{1}{ i \slashed{\partial} - m_a + \frac{\beta_1}{d_1} \varphi_1} , \;\;\;\;  & \mbox{for n $=$ 1},
\nonumber
\\
\frac{\varphi_n}{d_n} &= -  i  \frac{\varphi_1}{d_1}  \frac{\partial \beta_1}{\partial  \varphi_n}
Tr \frac{1}{ i \slashed{\partial} - m_a + \frac{\beta_1}{d_1} \varphi_1} , \;\;\;\;  & \mbox{for n $>$ 1},
\end{eqnarray}
where 
\begin{eqnarray} \label{deriv-2}
 \frac{\partial \beta_1}{\partial \varphi_2} &=& 
\frac{d_1}{\beta_1} \frac{\beta_2}{d_2} ,
\nonumber
\\
 \frac{\partial \beta_1}{\partial  \varphi_3} &=& 
\frac{1}{2 \beta_1} \frac{2 d_1 }{d_2}  \frac{\partial \beta_2}{ \partial \varphi_3} = 
\frac{d_1}{\beta_1}  \frac{\beta_3}{ d_3 }  \varphi_2,
\nonumber
\\
&...&
\nonumber
\\
 \frac{\partial \beta_1}{\partial  \varphi_n} 
&=& 
\left( \prod_2^{n-1} \frac{\varphi_i}{\beta_i}
\right) \frac{d_1}{d_n} \frac{\beta_n}{\beta_1} .
\end{eqnarray}

Therefore the gap equations can be written as:
\begin{eqnarray}   \label{gap-n-2}
\frac{\varphi_1}{d_1}  &=&  \frac{\beta_1}{d_1} 
I_{\Lambda}  , \;\;\;\;\;\;\;\;\;\;\;\;\;\;\;\;\;\;\;\;\;\;\;\;\;\;\;  \mbox{for n $=$ 1},
\nonumber
\\
\frac{\varphi_n}{d_n} &=&   \frac{\beta_n}{d_n \beta_1}
\left( \prod_{i=2}^{n-1} \frac{\varphi_i}{\beta_i}
\right) 
I_{\Lambda}  , \;\;\;\;   \mbox{for n $>$ 1},
\end{eqnarray}
Where the following quantity was defined:
\begin{eqnarray}
I_{\Lambda} =-  i Tr \frac{1}{ i \slashed{\partial} -m_a^*},
\end{eqnarray}
where the effective mass  is given by: $ m^*_a = m_a - \frac{\beta_1}{d_1} \varphi_1^{(0)}$ 
being  written in terms of the
vacuum expected value of the auxiliary  fields.   $\varphi_i^{(0)} = <\varphi_i>$.

Therefore for all the gap equations we  can write:
\begin{eqnarray}   \label{gap-n-3}
\frac{\varphi_n}{\beta_n} &=&   \frac{\varphi_1}{\beta_1}
\left( \prod_{i=1}^{n-1} \frac{\varphi_i}{\beta_i}
\right) 
,
\end{eqnarray}
This means that, for the model (\ref{series-I}), all the higher order 
condensates are factorized into   the lowest order
 condensate, $<\bpsi_a \psi_a>$. 
Therefore if the first gap equation has non trivial solution(s), all the  solutions for the others are obtained.
However, for this gap equation all  the set of coupled algebraic expressions 
(\ref{betas}) for $\beta_n$ must be solved together.
This system turns out to be highly non linear and complicated.
For these equations, 
all the variables and parameters  were rescaled by an arbitrary constant of
dimension of mass,
 $\mu$,
such that $g_{2^n} = \tilde{g}_{2^n} (\mu^{4- 3.2^n})$, $m_a=M=\tilde{M} \mu$, 
$\varphi_n=\tilde{\varphi}_n \mu^2$
and $\beta_n=\tilde{\beta}_n (\mu^{2- 3.2^{n-1}})$. 
Besides that, momentum is also rescaled by $k=\tilde{k} \mu$, and 
by performing the momentum integration with a covariant Euclidean cutoff $\Lambda$,
it  rescales to $\Lambda=\tilde{\Lambda} \mu$.
The  set of equations (\ref{betas}) and also (\ref{gap-n-1},\ref{gap-n-3}) become  independent of $\mu$.
For lower dimensions, it should appear a maximum number
 number of components $N_r$ for which the gap equations
(in particular for $\varphi_1$) provide non trivial results
whereas for higher dimensions this issue should be less restrictive.
In the next section this model 
will be expanded 
by considering three a.f., i.e. $N=3$.

In Table \ref{tab:tab1} there are  few
numerical   solutions for the case of
equal masses $\tilde{m}_a=\tilde{M}=0.9$, $N=3$ and $N_r=2$
in four dimensions  with $\tilde{\Lambda}=4$.
With $N_r=2$ it is possible to 
consider non trivial polynomial interactions up 
to $(\bpsi_a \psi_a)^{2^n=8}$, i.e. for $n=3$.
This regime of the phase diagram
 shows explicitely the validity of the minimal procedure,
i.e. within the weak field approximation,
since $\beta_2 \varphi_2$ 
can fluctuate smoothly  around the (positive or negative) vacuum value
provided $g_1 > |\beta_2 \varphi_2|$.
The same for $\beta_3 \varphi_3$ with respect to $g_4$.
Therefore expressions (\ref{betas}) have real solutions for weak fluctuations.
The same analysis applies for all $\beta_n$ ($n<N$).
It is interesting to note that,  the condensates go to zero 
for larger coupling constant $\tilde{g}_2$ because the
other coupling constants were kept constants.
It can be noted, analysing $\tilde\varphi_1^{(0)}$ for the third, fourth, fifth and sixth 
lines,
that to obtain the usual symmetric ground state  (where $M^*/M=1$),
the order in which the coupling constants are set to zero
might be  relevant.

\begin{table*}
\caption{  Approximated dimensionless parameters and
 resulting variables from equations (\ref{betas}) and (\ref{gap-n-1})
for $N_r=2$, $N=3$,
$\tilde{M}=0.9$ and 
$\tilde\Lambda=4$. $\tilde{\varphi}_2^{(0)}$ and $\tilde{\varphi_3}^{(0)}$
 are
unambiguosly 
determined by expressions (\ref{gap-n-3}).}
 \label{tab:tab1}
\begin{tabular*}{\textwidth}{@{\extracolsep{\fill}}lrrrrrrrl@{}}
\hline
 $\tilde{g}_2$  &  $\tilde{g}_4$  & $\tilde{g}_8$  &   $\tilde\beta_1$   &  
$\tilde\varphi_1^{(0)}$    &  $M^*/M$    & $\varphi_1^{(0)} = \tilde{\varphi}_1^{(0)} \beta_1$
  &  $\tilde{\beta}_2$    &    $\tilde{\beta}_3$
\\
\hline 
$10^{-5}$  & $10^{-5}$  & $10^{-5}$   &  0.0   & 0.60 & 1.00 &  0.0 & 0.004 &  0.004 
\\
0.0001  & 0.001 & 0.001  &  0.15  & 0.65 & 1.02 &  0.19 & 0.055 & 0.044  
\\
0.01  & 0.001 & 0.001  &  0.15  & 0.65 & 1.02 &  0.19   & 0.055 &  0.045 
\\
0.01  & 0.01 & 0.01  &  0.20   & 0.66 & 1.03 &  0.13  & 0.18 &  0.14  
\\ [0.5ex]
0.02  & 0.01 & 0.01  &  0.23   & 0.64 & 1.05  & 0.15  & 0.18 &  0.14 
\\
0.03  & 0.01 & 0.01  &  0.29   & 0.66 & 1.06 & 0.19 &  0.18 &  0.14 
\\
0.1  & 0.01 & 0.01  &  0.47  & 0.70 & 1.18 &  0.33  & 0.18  &  0.14
\\
1  & 0.01 & 0.01  &  1.46  & 1.02 & 3.41  &  1.49  &  0.25  &  0.14
\\
\hline
\end{tabular*}
\end{table*}

\subsection{ Expansion of the model }

In the following 
a large fermion mass  (zeroth order) derivative expansion
 of the determinant is done such as to write down an 
effective polynomial model for the scalar fields.
The fermion determinant can be written as:
\begin{eqnarray} \label{determ-I}
Tr \ln \left[ 1 + D_a  \left(  \beta_1 \frac{\varphi_1}{d_1}  
 \right)
\right] + Tr \ln D^{-1}_a,
\end{eqnarray}
where $D_a = \frac{1}{i \slashed{\partial } - m^*_a}$.
The first terms expansion corresponds to:
\begin{eqnarray} \label{deriv-exp}
&&  S_{eff} \simeq
S_{eff,(0)} [\varphi_i^{(0)}] 
+ \sum_i^N
\sum_j^N
 \frac{1}{n_i ! n_j !}
 \int_{x_1, x_2} 
  \left.
\frac{\delta^2 S_{eff}}{\delta \varphi_i (x_1) \delta  \varphi_j(x_2)} 
 \right|_{{\varphi_i}=\varphi_i^{(0)} }
\varphi_i (x_1) \varphi_j (x_2) 
+
h.o.
\nonumber
,
\end{eqnarray}
where $\int_{x_1, x_2} = \int d x_1 \int d x_2$, 
$h.o.$ 
stands for 
(even) higher order  derivatives, $n_i, n_j=0,1,2$
 are such that  $n_i+n_j=2$ .
The first derivative term
is set to zero due to the stability condition. 
A constant multiplicative  factor appears for each of the    derivative
and therefore 
a field redefinition can be done to simplify the resulting expressions.
These   field redefinitions are  given by:
\begin{eqnarray}
\varphi_1 \to \varphi_1 \frac{\beta_1}{d_1} \equiv \chi_1,
 \;\;\;\;\;
\varphi_2 \to \varphi_2 \left(
 \frac{\varphi_1^{(0)}}{\beta_1}  \frac{\beta_2}{d_2} \right) \equiv \chi_2,
\;\;\;\;
 \varphi_3 \to \varphi_3 \left(
 \frac{\varphi_1^{(0)}}{\beta_1}  \frac{\varphi_2^{(0)}}{\beta_2}
\frac{\beta_3}{d_3} \right) \equiv \chi_3 ,
\;\;\;\; ...
.
\end{eqnarray}

With this redefinition, all the auxiliary fields $\chi_i$ will have the same dimension.
By assembling the interaction terms,
 it yields for the first four auxiliary fields:
\begin{eqnarray} \label{eff-I}
{\cal V}^{eff}_I &=&
\frac{1}{2}  \left[ ( c_2 +c_{2,1})
\chi_1^2
+  \left( c_2 + 
c_{2,2}
\right)
\chi_2^2
 + 
 \left( c_2 
+ c_{2,3} 
\right)
\chi_3^2  \right]
 + \sum_{n\geq 3} 
\left[  c_{ n}  \chi_1^n +  ( c_n + c_{n,2} )\chi_2^n + 
\right.
\nonumber
\\
&& \left. 
 ( c_n + c_{n,2} + c_{n,3} )\chi_3^n  \right]
+ \sum_{i,j,k}
 t_{i,j,k}
\chi_1^i   \chi_2^j \chi_3^k 
,
\end{eqnarray}
where 
$t_{i,j,k}$ are defined for  
 $ i+ j + k = m \geq 2$ 
being 
$i,j,k=0,....m$
where at least two indices are different from zero,
and where $c_n$ and $c_{n,m}$ are the resulting self interaction coupling 
constants and contributions for masses, and also the 
couplings $t_{i,j,k}$ are those couplings between at least two different components,
being that at least two indices are non zero, i.e. $i,j\neq 0$ or $i,k \neq 0$ and so on.
$c_2$ are the terms provinient from the unity integrals of the 
auxiliary fields.

In the limit of same masses for all the fermion components ($m_a = m$)
  the same kernels $D_a=D$  are obtained.
Should the Lagrangian fermion masses of each of the components be
 different
there would be a degree of freedom more in the kernels $D \to D_a$ with which
the arguments below can be drawn although not necessarily 
yielding the same conclusions.
This allows the resulting coupling constants to be written  and defined 
in an uniform notation.
They can be written as:
\begin{eqnarray}
c_2 &=& Tr D^2, 
\;\;\;\;\;
c_{2,1} = \frac{d_1^2}{\beta_1^2}
\nonumber
\\
c_{2,2} &=& \frac{d_1}{\beta_1 {\varphi_1^{(0)}}} 
Tr D  + \frac{d_2 \beta_1^2}{\beta_2^2}
, 
\nonumber
\\
c_{2,3}  &=&  \frac{d_2 \beta_1}{\beta_2  {\varphi_1^{(0)}} {\varphi_2^{(0)}} }
 Tr  D + \frac{\beta_1^2 \beta_2^2 d_3}{{\varphi_1^{(0)}}^2 {\varphi_2^{(0)}}^2 \beta_3^2 },
\end{eqnarray}
The couplings $t_{i,j,k}$  are due to the parametric dependence of the 
term $\beta_1 \varphi_1$ of expression (\ref{determ-I})  on the other fields,
$\beta_1 \varphi_1 \to \beta_1[\varphi_2, \beta_2] 
\varphi_1 \to \beta_1 [\varphi_2, ... \varphi_N]
\varphi_1$.
The massive terms  of the auxiliary fields cannot  be equal 
according to the above results  however they might be 
nearly equal if the contributions $c_{2,n}$ are progressively smaller, 
i.e. for progressively  large $\beta_n$ for larger $n$:
($\beta_3^2 > \beta_2^2 > \beta_1^2$ )
and/or 
$|\varphi_{n}^{(0)}|<< |\varphi_{n+1}^{(0)}|$.
Although this limit could  be spoiled by the weak field condition 
observed for expressions (\ref{betas})
but it is not in all the cases analysed in this section.
Even if the field redefinition above is not done, the same limit is achieved for
the a.f. $\varphi_n$'s
the case in which $\beta_n^2/\beta_{n-1}^2$ is small what
is achieved nearly in the same regime as the progressively large condensates
regime.
 For the results of Table (\ref{tab:tab1})
the weakest coupling constants $g_2$   (largest condensate values)
correspond nearly  to values in which the coefficients $c_{2,n}$ become 
smaller and the  limit below of approximated symmetric effective 
potential is valid.
For the second order interactions between the different components $\chi_i \chi_j$,
given by the terms $t_{i,j,k}$ 
of expression (\ref{eff-I}),
it yields:
\begin{eqnarray} \label{symm}
&& t_{1,1,0} = t_{1,0,1}   = t_{0,1,1} - t'_2
,
\end{eqnarray}
where 
\begin{eqnarray}  \label{couplings-I3}
t_{1,1,0} &=& -  \frac{d_1}{\beta_1 \varphi_1^{(0)}} Tr  D_a + Tr D^2_a,
\nonumber
\\
t_2' &=&  \frac{d_2 \beta_1}{\varphi_1^{(0)}    \varphi_2^{(0)}} Tr D_a, 
\end{eqnarray}
All these couplings have the same dimension, in d=4 they have dimension mass square.
For progressively higher order interaction terms,  different structures appear for
higher order  auxiliary fields $\chi_4, \chi_5$ and so on.
In this case, as well as in higher order interactions, there is a privileged role of the 
first component $\varphi_1$ (now $\chi_1$) over the others gererating a sector  of  the model
of higher symmetry than
the full original  model.
This different role for the lowest order fields, in particular $\varphi_1$, is more apparent 
and explicitely in the case  mean fields are zero.
These second order terms might  become equal by adjusting the coupling constants 
of the original model, defined in expression (\ref{series-I}),
and consequently the parameters $\beta_i$ and $d_i$, 
such that it could yield instead:
$ t_{0,1,1,0} - t'_2= t_{0,1,0,1} -  t'_2 = 0$.
Also, in the same limit of progressively larger values of the condensates
$\varphi_{n}^{(0)} << \varphi_{n+1}^{(0)}$ mentioned above,
the differences in the coupling constants become smaller.

The expansion at  third order, for which ($i+j+k= 3)$ with at least two indices non zero), also
yields
 terms with identical coeficients and terms with slightly  different coefficients.
It can be written that:
\begin{eqnarray}
t_{2,1,0} &=&  t_{2,0,1}  = t_{1,2,0}  = t_{1,0,2} 
= t_{0,2,1} -  t_3 = t_{0,1,2} - t_3' ,
\end{eqnarray}
where $t_3 - t_3' \sim \varphi_1^{(0)}/\beta_1$,  which
 is small with respect to $t_{2,1,0}$ and other terms
in the limit of small  $\varphi_1^{(0)}/\beta_1$ considered above.
 It appears an approximated identification for all the couplings of the type:
\begin{eqnarray}
t_{2,1,0}  \sim t_{2,0,1}   \sim t_{0,2,1}  \sim t_{1,2,0} \sim t_{0,1,2}  ...
\end{eqnarray}
This is valid also for the masses in expression (\ref{eff-I}) and all the 
higher order couplings.
Finally the term $\varphi_1 \varphi_2 \varphi_3$ has a coefficient 
$t_{1,1,1}$ whose difference $t_{1,1,1} - t_{1,2,0} \sim \beta_n/\beta_{n-1}$
and $t_{1,1,1} - t_{1,2,0} \sim \varphi_{n-1}^{(0)}/\varphi_{n}^{(0)}$.
By considering that there is a factor $3$ with respect to the 
effective interactions of the type $\varphi_n^3$
 from the combinatorial factor in the expansion,
in this limit,   it yields in general the following
effective potential:
\begin{eqnarray} \label{Vlargecondens}
{\cal V}_{eff}^{large \;\; \varphi^{(0)}_i} \simeq 
\sum_{n=2} g_n \left( \chi_1 + \chi_2 + \chi_3 + ..
\right)^n .
\end{eqnarray}
The expansion also provides kinetic terms which appear by writing the 
kernels 
$D$ with a part diagonal in momentum space and another part diagonal 
in coordinate space \cite{mosel},
 i.e.
$
D =  - (i \slashed{\partial} + m^*) \cdot S_0,
$
where $S_0 = 1/(k^2 + {m^*}^2)$.
The lowest order derivative terms in the limit considered above 
for the expression (\ref{Vlargecondens}), is the following:
$$
\Delta {\cal L}_{eff} 
= \frac{F}{2} \partial_{\mu} (\chi_1 + \chi_2 + \chi_3 + ...) \partial^{\mu}
(\chi_1 + \chi_2 + \chi_3 + ...) ,
$$
 where $F$ is  a constant to be calculated from the expansion.
This effective boson  model is invariant under any transformation
that keeps the length $\left( \sum_i \chi_i \right)$ invariant.
One set of continuous  transformations is given by:
\begin{eqnarray}
\chi_1 \to \chi_1 + b_1 \chi_2 + c_1 \chi_3,
\nonumber
\\
\chi_2 \to \chi_2 + a_2 \chi_1  - c_1 \chi_3,
\nonumber
\\
\chi_3 \to \chi_3 - a_2  \chi_1 -  b_1 \chi_2.
\end{eqnarray}
For this transformation considering three fields, N=3,
there are three parameters in the transformations. For the case of $N$ fields
 there will have
$N^2-2N$ 
parameters/generators of the algebra.
The resulting particle excitations present therefore the same mass, since these expressions
 already  correspond to fluctuations dynamics.
The resulting algebra for this set of transformations
will not be discussed here.
Apart from the above symmetry this effective model is invariant 
under  simple permutations of the fields such as 
($\chi_1 \to \chi_2$, $\chi_2 \to \chi_3$ and $\chi_3 \to \chi_1$).

\subsection{ Removing  weak fields conditions}
\label{sec:no-weak}

Two  ways of overcoming the limitation of weak field are given in this Section.
Although they might require a non minimal procedure they yield the same result as shown above.
However one might also simply require the scalar fields $\varphi_n$,
 for $n<N$, to only assume positive
values in which case the dynamics would be resctricted to one side of the effective potential
$V_{eff} (\varphi_i >0)$.
 
The first more general solution 
is to  introduce  further set of  auxiliary fields $\xi_n$  with 
the same shifts provided above with different parameters $\beta_n'$.
In this case expression (\ref{betas}) can be rewritten as:
\begin{eqnarray}
\beta_n^2 + {\beta_n'}^2 &=& 
2 d_n \left( g_{2^n} + \frac{\beta_{n+1} }{d_{n+1} } \varphi_{n+1}
 + \frac{{\beta_{n+1}}' }{d_{n+1} } \xi_{n+1}
\right)    
 \;\;\;\;   \mbox{for n $<$ N},
\nonumber
\\
\beta_n^2 + {\beta_n'}^2 &=& 2 d_n g_{2^n}       
\;\;\;\;\;\;\;\;\;\;\;\;\;\;\;\;\;\;\;\;\;\;\;\;\;\;\;\;\;\;\;\;\;\;\;\;\;\;\;\;\;\;\;\;\;\;
 \mbox{for n $=$ N}.
\end{eqnarray}
It looks 
 the number of free parameter now is doubled. 
However to avoid limiting to weak field, 
$\beta_n'$ is associated to the imaginary part of the  functions $\beta_n$ 
being that the corresponding field
$\xi_n$ might  not  correspond  necessarily 
to a physical degree of freedom to avoid
 double counting.

A second,
 more general,
  solution to overcome the eventual
 limited range of values of the auxiliary fields
is to consider vector parameters for  the shifts 
in the Gaussian integrals (\ref{gaussian}).
For the sake of the argument, let us consider the four dimensional case
with $d_n=1$ to provide an example. 
The a.f. in the unity Gaussian integral can be written as:
\begin{eqnarray} 
- \frac{\varphi_1^2}{2} &\to& -\frac{1}{2 A^{\mu}A_{\mu}}
\left( A^\mu \varphi_1- C_1^\mu \bpsi \psi \right)
\left( A_\mu \varphi_1- C^1_\mu \bpsi \psi \right),
\nonumber
\\
- \frac{\varphi_2^2}{2} &\to&
-\frac{1}{2 A^{\mu}A_{\mu} }
\left( A^\mu  \varphi_2- C_2^\mu (\bpsi \psi)^2 \right)
\left( A_\mu  \varphi_2- C^2_\mu (\bpsi \psi)^2 \right),
\nonumber
\\
- \frac{\varphi_3^2}{2 } &\to& 
-\frac{1}{2  A^{\mu}A_{\mu}}
\left( A^\mu \varphi_3 - C_3^\mu (\bpsi \psi)^4 \right)
\left( A_\mu \varphi_3 - C^3_\mu (\bpsi \psi)^4 \right),
\end{eqnarray}
where $A^\mu$ is constant.
Only two non zero components are  enough, i.e.
$A_\mu = (A_0,A_1,0,0)$
and the reason is that it must have the minimum number 
of degree of freedom (arbitrary constants to be determined below)  and it should account 
for the possibility of non trivial contractions $C_\mu^n \cdot A^\mu$ (where $C_\mu^n$
have two components as well).
 In any case $A_\mu$  can be choosen so that: $A_0 = A_1 \sqrt{2}$
 and then it can be written in terms of only one free parameter
$A_1$.
 The vectors $C_\mu^n$ are parameters that might be functions of the auxiliary fields
in the same way the parameters $\beta_n$  of the minimal procedure  do.
Also it is enough to have  two non zero components, as shown below, to 
allow for positive and negative normalization $C^\mu_n \cdot  C_\mu^n$.
With these shifts the  cancelations are obtained with the following relations:
\begin{eqnarray} \label{vec-cancel}
\frac{C_3^\mu \cdot C_\mu^3}{A_\mu A^\mu} &=& 2 g_8,
\nonumber
\\
\frac{C_2^\mu \cdot  C_\mu^2}{A_\mu A^\mu} &=& 2  \left(
g_4 + \frac{C_\mu^3  \cdot A^\mu}{ A^{\mu} \cdot A_{\mu}} \varphi_3 \right),
\nonumber
\\
\frac{C_\mu^1  \cdot C^\mu_1}{A_\mu A^\mu} &=& 2 
\left(
g_2 + \frac{C_\mu^2  \cdot A^\mu }{ A^{\mu} \cdot A_{\mu}} \varphi_2
\right),
\end{eqnarray}
being that $N-1$ of these vector parameters $C_\mu$ 
become functions of 
some of the auxiliary fields (in the case of $N=3$ they are $C^\mu_2$ and $C_1^\mu$).
It yields an effective action with the same shape and structure of expression (\ref{Seff-n}), given by:
\begin{eqnarray} \label{vec-par}
S_{eff} &=& 
- i Tr \ln \left(
i \slashed{\partial} - m_a + g_1 + \frac{C_1^\mu A_\mu \varphi_1}{ A^{\mu}A_{\mu}} 
\right)
- \int_x \left( 
\frac{ \varphi_1^2}{2}  + \frac{ \varphi_2^2}{ 2 }  
+ \frac{\varphi_3^2}{2} 
\right).
\end{eqnarray}
The difficulty with this parameterization
 might be the number of free parameters ($A^\mu$ and $C_\mu^n$) that 
is larger than the number of expressions (\ref{vec-cancel}).
However this can be avoided
by   a direct  identification with the minimal procedure
which   can be  given
if:
\begin{eqnarray}
\beta_n^2 = C_\mu^n C^\mu_n,
\hspace{1cm}
 \mbox{and} \hspace{1cm}
A^\mu \cdot A_\mu = 1.
\end{eqnarray}
Now, the case of negative  $\beta_n^2$  corresponds to 
the negative normalization of the vector $C_n^\mu C_\mu^n$,
consequently  $\varphi_n$ ($\chi_n$) do not need to assume complex values.
Therefore  further choices for  the parameters can be done, such as:
$C_3^\mu=(L_0,0,0,0)$
for positive $g_8$,
$A^\mu=(\sqrt{2},1,0,0)$ for the normalization of the Gaussian integrals, 
 $C_2^\mu = (K_2,D_2 ,0,0)$ 
and $C_1^\mu=(K_1,D_1,0,0)$.
With this, the second and third expressions of (\ref{vec-cancel})
could be expected to fix four undetermined parameters ($K_1, K_2, D_1$ and $D_2$).
The only choice that makes these expressions non ambiguous is 
that $K_i = 0$ and $D_i \neq 0$  if $C_\mu^i C^\mu_i < 0$ and
 $K_i \neq 0$ and $D_i = 0$  if $C_\mu^i C^\mu_i  \geq 0$.
Therefore to  eliminate the ambiguity in defining the components of $C_\mu^n$:
$K_1$ and $K_2$
 become  functions or parameters
 that parameterize  only the
positive values of  $C_\mu^n C^\mu_n$, and 
$D_1$  and $D_2$ are functions or parameters
that parameterize  only the 
 negative values of $C_\mu^n C^\mu_n$.
Second and third conditions (\ref{vec-cancel}) can then be written 
in the two cases of positive or negative arguments as:
\begin{eqnarray} \label{heaviside}
K_i^2  =  2 [g_{2^i} + ( K^{i+1} \sqrt{2}- D^{i+1} ) \varphi_{i+1} ] \; \geq 0,
\nonumber
\\
- D_i^2   =  2 [g_{2^i} + ( K^{i+1} \sqrt{2}- D^{i+1} ) \varphi_{i+1} ] \; < 0.
\end{eqnarray}

\section{More general series $\sum_n (\bpsi_a \psi_a)^{n}$  }
\label{sec:geral}

 In this section 
we consider a more general series of billinears.
The generating functional   will be given by:
\begin{eqnarray} \label{Z-L-original-2}
Z =  \int {\cal D} [\bpsi_a, \psi_a]
e^{  \; i \int_x {\cal L [\psi_a, \bpsi_a]} },
\end{eqnarray}
where a general  series of   bilinears of the following form will be considered:
\begin{eqnarray} \label{series-II}
{\cal L} =
\bpsi_a (x) \left( i \slashed{\partial} - m_0 \right) \psi_a (x)
+ \sum_{n=2}^{N}  g_{2n} \left( \bpsi_a \psi_a \right)^n ,
\end{eqnarray}
 where  the case of even $N$  will be addressed,
and the case for $N$ odd 
will be discussed below shortly.
In
 each bilinear there is an implicit
 sum over $a$ and the mass term is therefore diagonal.

The auxiliary fields are introduce by means of
 $N/2$  unity integrals that are given by:
\begin{eqnarray}
N'  \int \; {\cal D} [\xi_m]
 e^{- i \int_x \sum_m^{N/2} \frac{1}{2 d_m} \xi^2_m (x)} 
 = 1,
\end{eqnarray}
where $d_m$ are constants, eventually they can be set to $1$.

The simplest necessary shifts of the auxiliary fields that cancel out the interactions  
can be written as:
\begin{eqnarray} \label{shifts-n-1}
\frac{1}{2 d_m} \xi_m^2 &\to&
\frac{1}{2 d_m} \left(   \xi_m -   B_m (\bpsi_a \psi_a)^{m}
 -   A_m  (\bpsi_a \psi_a)^{m-1}
\right)^2  .
\end{eqnarray}
There are other possible shifts in the auxiliary fields, however 
these are the simplest ones that cancel out all the polynomial interactions.

The  cancelation of all the interactions occur if the  following relations hold:
\begin{eqnarray} \label{cancelations-even-ger}
g_4 &=& \frac{B_1^2}{2 d_1} + \frac{A^2_2}{2 d_2} 
- \frac{\xi_3 A_3}{d_3} - \frac{\xi_2 B_2}{d_2} 
,
\nonumber
\\
g_6 &=& \frac{A_2 B_2}{d_2}  - \frac{B_3}{d_3} \xi_3 
 - \frac{A_4}{d_4} \xi_4 ,
\nonumber
\\
g_8 &=& \frac{B^2_2}{2 d_2} + \frac{A_3^2}{2 d_3}
 - \frac{B_4}{d_4} \xi_4 ,
\nonumber
\\
g_{10} &=&  \frac{B_3 A_3}{d_3} 
,
\nonumber
\\
g_{12} &=&  \frac{B_3^2}{2 d_3} 
 + \frac{A_4^2}{2 d_4} 
\nonumber
\\
g_{14} &=& \frac{B_4 A_4}{d_4}
\nonumber
\\
g_{16} &=& \frac{B_4^2}{2 d_4}
\nonumber
\\
&& ...
\nonumber
\\
g_{2n} &=& \frac{B^2_{n/2}}{2 d_{n/2}} + \frac{A_{(n+2)/2}^2}{2 d_{(n+2)/2}} 
 - \frac{B_n }{d_n} \xi_n - \frac{A_{n+1}}{d_{n+1}}  \xi_{n+1}
\nonumber
\\
&& \;\;\;\;\;\;\; \mbox{n even, $n \leq N-1$} .
\end{eqnarray}
In particular for $n=N$
\begin{eqnarray}
g_{2N} = \frac{B_{N/2}}{2 d_{N/2}}.
\end{eqnarray}
All the discussion and remarks made in the last section applies here for the 
case of enforcing the weak field conditions or to lift them.

Since one of the aims of this calculation is to show the structure of the resulting
model for auxiliary fields, and to compare with the  model from the last section,
the series will stop at $N=6$, i.e. $(\bpsi \psi)^6$, such that 3 auxiliary fields
are needed.
According to the expressions above,  the parameters of the shifts $A_n,B_n$
 must  be considered to 
be field dependent. 
This dependence has an unique possible choice which is given by:
\begin{eqnarray} \label{cancelationsqq}
B_1 [\xi_2, \xi_3] &=& \sqrt{ (2 d_1)  \left(
g_4 -  \frac{A^2_2}{2 d_2} 
+ \frac{\xi_3 A_3}{d_3} + \frac{\xi_2 B_2}{d_2}
\right) }
\nonumber
\\
A_2  [\xi_3, \xi_4] &=&  \frac{d_2}{B_2}
 \left( g_6   + \frac{B_3}{d_3} \xi_3 
\right)  ,
\nonumber
\\
 B_2 [\xi_4, \xi_5] &=&  \sqrt{  (2 d_2)
\left(  g_8 - \frac{A_3^2}{2 d_3} 
 \right) 
},
\nonumber
\\
 A_3 [\xi_5, \xi_6] &=&   \frac{d_3}{B_3 }
g_{10}  ,
\nonumber
\\
B_3 [ \xi_6, \xi_7]  &=&   \sqrt{ 2 d_3
 g_{12} 
}.
\end{eqnarray}
In general, for $N$ even:
\begin{eqnarray}  \label{cancelations-gen}
&& 
B_{N/2} = \sqrt{g_{2n}   2 d_{n/2}} 
\;\;\;\;\;\;\;\; \mbox{$n=N$ even} ,
\nonumber
\\
&& B_{n}  = \sqrt{  (2 d_{n} )
\left( g_{4n} + \Xi
\right) }, 
 \;\;\;\; \mbox{ $n\leq N-1$ even} ,
\nonumber
\\
&& A_{n}   =
\frac{ d_{n} }{B_{n}}
\left( g_{2(n+1)} 
 + \frac{B_{n+1} }{d_{n+1}} \xi_{n+1} 
+\frac{A_{n+2}}{d_{n+2}}  \xi_{n+2}
\right)
\;\;\;\; n \geq 2 ,
\end{eqnarray}
where 
$$
\Xi = - \frac{A_{n+1}^2}{2 d_{n+1}} 
 + \frac{B_{n+1} }{d_{n+1}} \xi_{n+1} + \frac{A_{n+2}}{d_{n+2}}  \xi_{n+2}
$$
and therefore:
$
B_{n} = B_{n} [\xi_n, \xi_{n+1} , \xi_{n+2}]
$.
The limit of weak field is also assumed and the corresponding discussion
of the last section applies here.

The action can be rewritten as:
\begin{eqnarray}
S_{eff} &=& \int_x   \left[ 
\bpsi_a \left( i \slashed{\partial} -
\left( \tilde{M}
 \right)
\right) \psi_a 
 -
 \sum_{m}^{N/2} \frac{1}{2 d_m} \xi^2_m (x) - \frac{A_1( A_1 - 2 \xi_1)}{2 d_1} 
\right],
\end{eqnarray}
where 
$$
\tilde{M} = m_a - \frac{(B_1 \xi_1 - B_1 A_1)}{d_1} 
- \frac{A_2}{d_2} \xi_2
$$
Therefore $A_1,B_1$ encode  all the non linearities of the model.
The saddle point equations for this model 
 provides  relations between the 
ground state average of the auxiliary fields $\xi_n$ and 
the progressively large power of bilinears: 
$< ( \bpsi_a \psi_a )^n>$.
In the same way it was done in last Section, with  expressions 
(\ref{def-cond-I}),  
one has the following relation between the ground state averaged value of the 
a.f.  and the composite fermion condensates:
\begin{eqnarray} \label{def-cond-II}
\frac{\xi_n^{(0)} }{B_n} 
\equiv < (\bpsi_a \psi_a)^n > + \frac{A_n}{B_n} <(\bpsi_a \psi_a)^{n-1}>.
\end{eqnarray}
 
With the integration of fermions the remaining terms, neglecting an irrelevant constant, 
 are the following:
\begin{eqnarray}
S_{eff} &=& -   i \; Tr \ln \left( i  \slashed{\partial} -  
 m_a + \frac{(B_1 \xi_1 + B_1 A_1)}{d_1} 
+ \frac{A_2}{d_2} \xi_2
\right) 
-
 \int_x 
\sum_{m}^{N/2}  \frac{ \xi^2_m (x) }{2 d_m}.
\end{eqnarray}
where $C_1= \frac{A_1( A_1 - 2 \xi_1)}{2 d_1}$.

The gap equations for the homogeneous a.f.  are therefore the following:
\begin{eqnarray}
\frac{\xi_1}{d_1} &=&    \frac{ A_1}{d_1}  - \frac{B_1}{d_1} 
 \;   i \;  Tr \frac{1}{i \slashed{\partial} - m_a^*} ,
\nonumber
\\
\frac{\xi_2}{d_2} &=&     -  \left( 
\frac{A_2}{d_2} + \frac{\xi_1}{d_1} \frac{d_1 B_2}{B_1 d_2 }
\right) \;   i \;  Tr \frac{1}{i \slashed{\partial} - m_a^*}  ,
\nonumber
\\
\frac{\xi_3}{d_3} &=&     \left[
\left(  \frac{\xi_1}{d_1} +  \frac{ A_1}{d_1} \right)
  \frac{2 d_1}{2 B_1}  
\left(
\frac{A_3}{d_3}  - \frac{d_2 B_3}{ B_2 d_3}
 \frac{A_2}{d_2}  
  \right)   
- \frac{\xi_2}{d_2} \frac{d_2 B_3}{B_2 d_3}
\right] 
 i \;  Tr \frac{1}{i \slashed{\partial} - m_a^*} ,
\end{eqnarray}
where 
$$
I_\Lambda =  -  i \;  Tr \frac{1}{i \slashed{\partial} - m^*_a}
$$
and 
where the effective mass has been defined as:
\begin{eqnarray}
m^* = m_a - \frac{(B_1 \xi_1^{(0)} - B_1 A_1)}{d_1} 
- \frac{A_2}{d_2} \xi_2^{(0)} .
\end{eqnarray}
In this expression, the fields are   the homogeneous, mean field, solutions
of the gap equations.
Although the effective mass depends explicitely only on the first two a.f.,
the parameters $A_1, B_1, A_2$ depend  on the higher order a.f. as shown 
in expressions (\ref{cancelationsqq}-\ref{cancelations-gen}).
At the end, all of the a.f. $\xi_n$ contribute for the effective mass.
The first three  gap equations can be rewritten as:
\begin{eqnarray}
\frac{\xi_1^{(0)}}{B_1} &=&   \frac{ A_1}{B_1}  +   I_\Lambda,
\nonumber
\\
\frac{\xi_2^{(0)}}{B_2} &=&     \left( \frac{A_2}{B_2} + \frac{\xi_1^{(0)}}{B_1} 
\right) \; \left( -  \frac{ A_1}{B_1}  + \frac{\xi_1^{(0)}}{B_1} \right),
\nonumber
\\
\frac{\xi_3^{(0)}}{B_3} &=&     
 \left[
\left(  \frac{\xi_1^{(0)}}{  B_1} -    \frac{  A_1 }{  B_1}\right)
\left(
\frac{A_3}{B_3}  - \frac{ A_2  
 }{ B_2}
  \right)   
- \frac{\xi_2 }{B_2 }
\right]
  \left( -  \frac{ A_1}{B_1}  + \frac{\xi_1^{(0)}}{B_1} \right),
\end{eqnarray}
with the corresponding definitions of the parameters (functions) $A_i$ and $B_i$.
By  writing these expressions in terms of 
all the higher order fermion condensates,
for the general case $A_1 \neq 0$, one obtains:
\begin{eqnarray}
 <(\bpsi_a \psi_a)^{2}> &=& (<\bpsi_a \psi_a>)^{2}
-  \frac{A_1}{B_1} \left[   3 < \bpsi_a \psi_a > 
 + 2  \left(  \frac{ A_2}{B_2}  - \frac{A_1}{B-1} \right)
\right],
\nonumber
\\
  <(\bpsi_a \psi_a)^{3}> &=& 
<\bpsi_a \psi_a> 
\left[
 \frac{A_3}{B_3}  \left(  <\bpsi_a \psi_a >  + \frac{A_1}{B_1}  \right) 
 + <(\bpsi_a \psi_a)^2> -  \frac{A_1 A_2}{B_1 B_2}
\right] 
\nonumber
\\
&& 
- \frac{A_3}{B_3} <(\bpsi_a \psi_a)^2>
.
\end{eqnarray}
From these expressions we conclude that  if $A_1 =  0$ there is a complete factorization
of higher order condensates into the lowest order fermion-antifermion, i.e.:
\begin{eqnarray}
  <(\bpsi_a \psi_a)^{n}> =  (<\bpsi_a \psi_a>)^{n}.
\end{eqnarray}
The case in which $A_1 \neq 0$ corresponds to a constant shift of the first auxiliary variable,
$\xi_1$, which is 
associated to the lowest order condensate $<\bpsi \psi>$ and therefore to its
redefinition.

For $N$  odd, two of  the shifts above (\ref{shifts-n-1}) 
would receive contribution
another term.
The shifts for these two higher order  auxiliary fields (i.e. $(N-1)/2$ and $(N-1)/2 -1$,
being now the highest order a.f. is $(N-1)/2$)
must be modified to the following:
\begin{eqnarray} \label{shifts-n-2}
\frac{\xi_{(N-1)/2}^2}{d_{(N-1)/2} }  &\to &
\frac{1}{ d_{(N-1)/2} } 
 \left(
 \xi_{(N-1)/2} -   B_{(N-1)/2} (\bpsi_a \psi_a)^{(N-1)/2} \right.
\nonumber
\\
&& 
\left.
 -   
A_{(N-1)/2}  (\bpsi_a \psi_a)^{(N-1)/2-1} 
 -  C_{(N-1)/2}  (\bpsi_a \psi_a)^{(N-1)/2+1}
\right)^2 , 
\nonumber
\\
 \frac{\xi_{(N-1)/2 -1}^2}{ d_{(N-1)/2 -1}}  &\to&
\frac{1}{d_{(N-1)/2-1}} \left(  \xi_{(N-1)/2-1} 
-   B_{(N-1)/2-1} (\bpsi_a \psi_a)^{(N-1)/2-1}
\right.
\nonumber
\\
&& 
\left.
 -   
A_{(N-1)/2-1}  (\bpsi_a \psi_a)^{\frac{N-1}{2}-2} 
 -  C'_{(N-1)/2}  (\bpsi_a \psi_a)^{\frac{N-1}{2}+1}
\right)^2 .
\end{eqnarray}
The higher interaction term from these shifts,  $(\bpsi_a \psi_a)^{N+1}$, now requires a further
trivial cancelation 
relation, being that all the subsequent development is unchanged.

\subsection{ Expansion of the model }

In the following,
the same large fermion mass  expansion
 of the determinant of the last Section  is done such as to write down an 
effective polynomial model for the scalar fields.
For the case $A_1= 0$, the contribution of the fermion determinant  can be written as:
$
Tr \ln \left[ 1 + D  \left(  B_1 \frac{\xi_1}{d_1} + A_2 \frac{\xi_2}{d_2}
 \right)
\right] + Tr \ln D^{-1},
$
where $D = \frac{1}{i \slashed{\partial } - M^*}$.
The lowest order terms of the zero order derivative
 expanded model 
for the auxiliary fields can be written
as:
\begin{eqnarray} \label{masses-II}
&&  {\cal V}_{eff}^{(2)} = \frac{1}{2}
\left[ -\frac{1}{d_1} + i Tr D^2 \frac{B_1^2}{d_1^2}
\right]_{\xi_i=\xi_i^{(0)}} \xi_1^2 
-
\frac{1}{2} \left[
 \frac{1}{d_2}  -
 i Tr D^2\left( \frac{ \xi_1^{(0)} B_2 }{ B_1 d_2} \right)^2 
+ i Tr D \frac{\xi_1^{(0)} B_2^2 d_1 }{B_1^3 d_2^2} 
\right]_{\xi_i=\xi_i^{(0)}} \xi_2^2
\nonumber
\\
&-&  
\frac{1}{2} \left[
 \frac{1}{d_3}  
+ \frac{i \delta^2 }{\delta \xi_2^2}
 Tr \log \left( i \slashed{\partial} + \tilde{M}
\right)
\right]_{\xi_i=\xi_i^{(0)}} \xi_3^2
+ \sum_{i \neq j}
c_{i,j}    \xi_i (x)  \xi_j (x)
+ \sum_{n_1, n_2. n_3}
c_{n_1,n_2,n_3}    \xi_1^{n_1} (x)  \xi_2^{n_2}  (x)  \xi_3^{n_3} (x)
 ,
\nonumber
\\
&&  \;\;\;\;\;\; \mbox{
($n_1 + n_2 + n_3 \geq 3$)}
,
\end{eqnarray}
where the second order term for $\xi_3$ was not written explicitely 
because its expression 
is quite long,
and it  does not really bring relevant information for the discussion below.
This resulting effective potential has a lower degree of symmetry than
the one 
derived in  Section II,
given by  expression (\ref{eff-I}).
The fields can  be redefined,
in the way it was done in the last section,
i.e. by means of $\xi_i \to \omega_i   G[\xi_i^{(0)},A_i,B_i]$ with convenient choice 
of the factors
such as to obtain  an unique mass term, 
$
\frac{m^2}{2} ( \omega_1^2 + \omega_2^2 + \omega_3^2  +..)
$.
However,
 the remaining interactions  will have a much   lower level of symmetry,
i.e. $c_{1,2} \neq c_{1,3} \neq c_{2,3} ...$, or 
 $c_{3,0,0} \neq c_{0,3,0} ... $ or $c_{2,2,0} \neq c_{2,0,2} ... $ and so on. 

Contrarily to the case analyzed in the  Section II, the limit of very large
condensates, does not yield an effective potential with any apparent symmetry.
If one considers the limit of zero condensates,
one reaches a non trivial model for the fields $\xi_1$ and $\xi_2$
 only, 
independently of the number of auxiliary fields considered.
Even in this case  it does not have
 any  apparent symmetry.
It can be written as:
$
V_{eff} = 
c_{2,1} \xi_1^2 + c_{2,2} \xi_2^2 + c_{12}  \xi_1 \xi_2 + 
c_{3,1} \xi_1^3 + c_{3,2} \xi^3_2 + ... $,
without an usual and satisfactory relation between the resulting 
masses and effective 
coupling constants.

Given the  two different resulting boson effective models found 
in Sections II and III, 
it must be noted 
 that the series presented in the previous section, expression (\ref{series-I}),
corresponds to a particular case of the more general series given in expression (\ref{series-II}).
The procedures adopted in both cases were the same and they are based  in the  introduction
of
the least number of auxiliary fields with the corresponding shifts.
Therefore, by choosing particular terms of the most general series with
particular values of the coupling constants, one might derive
 a secondary level boson effective 
model with a higher or lower degree of symmetry
very close to the ground state given by a strong enough mean field.

\section{ Series of interactions   $\sum_n (\bpsi_a \gamma_{\mu} \psi_a
\cdot 
\bpsi_b \gamma^{\mu} \psi_b)^{n}$}
\label{sec:vector}

The  
 local limit of an effective fermion model for the case 
of vector field exchange
can be written as:
\begin{eqnarray} \label{series-III}
{\cal L} =
\bpsi_a \left( i \slashed{\partial} - m_a \right) \psi_a 
+ \sum_n^{N}  g_{2n} 
(\bpsi_a \gamma_{\mu} \psi_a \cdot 
\bpsi_b \gamma^{\mu} \psi_b)^{n} , 
\end{eqnarray}
where $g_{2n}$ are the effective coupling constants with 
dimension: $[g_{2n}] = M^{-d+2n}$, where $M$ has dimension of mass,
$m_a$ are the masses for each of the fermion species
and the index 
$a,b=1...N_r$ stands for the fermion components.
In
 each bilinear there is a sum over $a, b$ 
and the mass term is therefore diagonal.

The auxiliary  fields  will be introduced by means of the  following unity
 integrals  multiplying the generating functional:
\begin{eqnarray} \label{gaussian}
N'  \int \; {\cal D} [\varphi_n]
 e^{- \frac{i}{2}  \int_x \left( \sum_2^{N}   \varphi^2_n (x)
+ \varphi_\mu \varphi^\mu
\right) } 
 = 1 ,
\end{eqnarray} 
where $N'$ is a normalization constant.
The necessary shifts of the a.f. needed to cancel out all the interactions 
can be made minimal shifts, i.e.,  the simplest shifts for the minimum number of
auxiliar fields which do not introduce Lagrangian terms  that  were not presented
in the original model. 
For the model of expression (\ref{series-I}) the shifts  are given by:
\begin{eqnarray} \label{shifts-4n-1}
\varphi_\mu^2 &\to& (\varphi_\mu - \beta_1 (\bpsi_a  \gamma_\mu \psi_a)
)^2,
\nonumber
\\
\varphi_2^2 &\to& \left(\varphi_2 - \beta_2 [ \bpsi_a \gamma_\mu \psi_a
\cdot \bpsi_b \gamma^\mu \psi_b ] 
\right)^2 ,
\nonumber
\\
\varphi_{2m}^2 &\to& \left(\varphi_{2m} - \beta_{2m} [  \bpsi_a \gamma_\mu \psi_a
\cdot \bpsi_b \gamma^\mu \psi_b ]^{m} 
- \alpha_{2m}  [ \bpsi_a \gamma_\mu \psi_a 
\cdot \bpsi_b \gamma^\mu \psi_b ]^{m-1}
\right)^2 \;\; (m \geq 2),
\end{eqnarray}
where $\beta_m$ and $\alpha_m$ are dimensionful parameters that are determined 
by imposing the corresponding cancelations of all polynomial  interactions.
Differently from the shifts of Section 2 here there are terms proportional
to $\alpha_{2m}$ which should not appear in the shifts of Section 2
to avoid the appearance of non existing terms in the original fermion
interactions.
The a.f. $\varphi_\mu$, $\varphi_2$ and  $\varphi_4$ represent therefore
vector fermion-antifermion, two fermion-two antifermion (four fermion) and 
eight fermion states respectively.

By considering the first four terms  ($N=4$ and $m=2$) in the potential (\ref{series-III}),
the conditions for the cancelations of the polynomial interactions
are given by:
\begin{eqnarray} \label{param-vec}
g_{8} &=& \frac{\beta^{2}_{4}}{2 } ,
\nonumber
\\
g_{6} &=&   \beta_4 \alpha_4 ,
\nonumber
\\
g_4 &=& -  \beta_4 \varphi_4
+ \frac{\beta_2^2 + \alpha_4^2}{2} ,
\nonumber
\\
g_2 &=& - \beta_2 \varphi_2 - \alpha_4 \varphi_4
+ \frac{\beta_1^2}{2}  .
\end{eqnarray}
If
 the parameters $\beta_n$ might 
 then considered to be functions of different a.f.
with the above conditions 
provided one  garantees that the shifts of the a.f. still have
unity Jacobian. 
In fact all these shifts yield     unity Jacobian
and different shifts that could cancel out the 
fermion interactions would  introduce
other non linearities and the need of extra a.f. or 
non unity Jacobians.
From the relations (\ref{param-vec}) all the free parameters 
are fixed unambigously.
They are given by:
\begin{eqnarray} \label{param-vec-II}
\beta_4 &=& \sqrt{2 g_8}, \;\;\;\;
\alpha_4 = \frac{g_6}{\beta_4},
\;\;\;
\beta_2 =  \sqrt{2 ( g_4 + \beta_4 \varphi_4 - 2 \alpha_4^2)}
\nonumber
\\
\beta_1 &=&   \sqrt{2 ( g_2 + \beta_2 \varphi_2 +  \alpha_4 \varphi_4)}
\end{eqnarray}
 From here on, it will be assumed that
the coupling constants values are such that there are well defined
 solutions for $\beta_n$ according to the discussion above
in Section 2.2.
By integrating out fermions it yields  the following effective action:
\begin{eqnarray} \label{Seff-vec}
S_{eff} &=& -  i Tr \log \left( i \slashed{\partial} - m_a 
+ \beta_1 \varphi_\mu \gamma^\mu
\right) 
-  \int_x ( \sum_{n=2}^N \frac{\varphi^2_n}{2}
+ \frac{\varphi_\mu \varphi^\mu}{2}),
\end{eqnarray}
where $Tr$  is the traces taken over all the internal indices and integration over space-time.
According to expressions (\ref{param-vec-II}), 
$\beta_1$ depends on all the fields $\varphi_n$ ($n \geq 2$), i.e.:
\begin{eqnarray}
 \beta_1  = \beta_1 [ \varphi_2, \beta_2 ]
= \beta_1 [\varphi_2, \beta_2 [\beta_3 [...[\beta_N]]]]
\end{eqnarray}
Therefore $\beta_1$ carries  the non linearities of the model.

The gap equations are given by:
\begin{eqnarray}
\varphi_\mu &=&    
  - 
\;   i \;  Tr \frac{2 \beta_1 \gamma_\mu}{i \slashed{\partial} - m_a + \beta_1 \gamma \cdot  \varphi } ,
\nonumber
\\
\varphi_2 &=&   -   i \;  Tr \frac{2 (\partial \beta_1/\partial \varphi_2) \varphi^\mu \gamma_\mu}{i
 \slashed{\partial} - m_a + \beta_1 \gamma \cdot  \varphi }  ,
\nonumber
\\
\varphi_4 &=&   -   i \;  Tr \frac{2 (\partial \beta_1/\partial \varphi_4) \varphi^\mu \gamma_\mu}{i
 \slashed{\partial} - m_a + \beta_1 \gamma \cdot  \varphi } .
\end{eqnarray}
where
\begin{eqnarray}
\frac{\partial \beta_1}{\partial \varphi_2} = \frac{\beta_2}{\beta_1} ,
\;\;\;\;\;\;
\frac{\partial \beta_1}{\partial \varphi_4} = \frac{\alpha_4}{\beta_1} -
\frac{\varphi_2 \beta_4}{\beta_1^3 \beta_2} 
\end{eqnarray}
It will be considered
  the vector a.f. does not develop
a non zero expected value in the vacuum, i.e. the solution for the first gap 
equation is trivially zero, $<\varphi_\mu> \to 0$.
This yields necessarily the trivial solutions for the other gap equations.

By expanding the effective action 
(\ref{Seff-vec}) around the minimum within 
a zero order derivative expansion a complicated structure appears
for the interaction between $\varphi_2$ and $\varphi_4$.
However, an interesting result is obtained 
in the case 
that 
$$
\frac{\alpha_4}{\beta_1} >> \frac{\varphi_2 \beta_4}{\beta_1 \beta_2}
$$
for which one may consider:
$
\varphi_2 \beta_4 << \beta_1 \beta_2
$.
This  means weak field $\varphi_2$ and $g_8 << g_2 g_4$,
and therefore 
$\partial \beta_1/ \partial \varphi_4 \sim \alpha_4/\beta_1$.
In this limit it is possible to write down 
an interesting form of the effective potential, it yields:
\begin{eqnarray} \label{eff-vec}
{\cal V}^{eff}_I &=& 
\frac{1}{2}
  \left[  
\varphi_\mu^2
\left( c_2 + 
c_{2,1}
\right)
 + c_2 \varphi_2^2 + c_2   \varphi_4^2 
\right]
+ V(\varphi_\mu)
+ \sum_{n=2}^N  \sum_m D_{n,m}
 (\varphi \cdot \gamma)^{m} \left(
 \frac{\beta_2}{\beta_1} \varphi_2 + 
\frac{\alpha_4}{\beta_1} \varphi_4 
\right)^n
\nonumber
\end{eqnarray}
where 
$c_2 = 1$,
$V(\varphi_\mu)$ depends exclusively on $\varphi_\mu$ 
which
 will not be analysed
here, $D_{n,m}$
 are the coefficients of each of the terms
of the expansion.
The coefficients can be calculated 
by considering the following quantity
$S_0 = \frac{1}{\gamma \cdot k - m_a}$,
although their explicit shape 
 do not bring any relevant information and 
thus they are not written
explicitely. 
Next 
 the following redefinition of the a.f. can be considered:
\begin{eqnarray}
\varphi_2  \to \frac{\beta_2}{\beta_1^3}  
\varphi_2 \equiv  \phi_2,
&\;\;\;&
\varphi_4 \to \frac{\alpha_4}{\beta_1^3} 
\varphi_4  \equiv  \phi_4,
\nonumber
\\
&&
\varphi_\mu \to  \beta_1 \varphi_\mu \equiv \phi_\mu.
\end{eqnarray}
It yields the following effective potential:
\begin{eqnarray} \label{Vefff}
{\cal V}_{eff} &=&  \frac{m_2^2}{2} \phi_2^2 +  \frac{m_4^2}{2} \phi_4^2
+ \frac{m^2_1}{2} \phi_\mu^2 +  V(\varphi_\mu)
+  \sum_{m=1} \sum_{n=1} 
 Tr 
\left(  
d_{n,m}
\right) 
(\gamma_\mu \cdot \phi^\mu )^{m}
\left( \phi_2 + \phi_4 \right)^n
,
\nonumber
\end{eqnarray}
where:
\begin{eqnarray} \label{masses}
m^2_2 = 
\frac{\beta_1^6}{\beta_2^2},
\;\;\;\;\;
m_4^2=  \frac{\beta_1^6}{\alpha_4^2},
\;\;\;\;\;
m_1^2 =
\left( \frac{1}{\beta_1^2} -  Tr S_0^2 \gamma_\mu \gamma_\nu \right).
\end{eqnarray}
 and 
where $d_{n,m}$
 are the coefficients
 from the expansion.

The important point here is that it appears 
an approximate
  symmetry for  two a.f. $\phi_2$ and $\phi_4$
(\ref{eff-vec}).
 For instance, if the masses  $m_2^2, m_4^2$ 
are neglected or very small, 
the remaining  part of the effective potential 
(\ref{Vefff}) is invariant under
the following transformations:
\begin{eqnarray}
\phi_2 &\to& a_2 \phi_2 + a_4 \phi_4 + a_0,
\nonumber
\\
\phi_4 &\to& b_2 \phi_2 + b_4 \phi_4 - a_0,
\end{eqnarray}
being that 
$$
a_2 + b_2 = 
a_4 + b_4 = 1 .
$$
If one includes the mass terms above,
the following 
  transformations  preserve the 
effective potential:
\begin{eqnarray}
\phi_2 \to  a_2 \phi_2 + a_4 \phi_4 ,
\;\;\;\;\;\;\;\;
\phi_4 \to b_2 \phi_2 + b_4 \phi_4 ,
\end{eqnarray}
being that  the following conditions must imposed:
\begin{eqnarray}
 b_2^2 &=& \frac{m_2^2}{m_4^2} (1 - a_2^2 ) ,
 \;\;\;\; \mbox{and} 
\;\;\;\; a_4^2 =  \frac{m_4^2}{m_2^2} (1- b_4^2) .
\end{eqnarray}
These expressions yield:
\begin{eqnarray}
b_4 = \frac{a_2 \frac{m_4^2}{m_2^2}}{
\sqrt{1 + a_2^2 \left(1+\frac{m_4^4}{m_2^4}\right)}},
\;\;\;\; \mbox{and} \;\;\;\;
\frac{m_2^2}{m_4^2} \simeq  \frac{g_6^2}{4 g_8 g_4},
\end{eqnarray}
where it has been considered only 
 the leading order for $\beta_2^2$ (i.e. 
for $\beta_2^2 \sim 2 g_4$) and $\beta_1^2 \sim 2 g_2$.
in the expression for $m_2^2/m_4^2$
with the masses given in (\ref{masses}).

It is interesting to note that 
the bosonized effective model (\ref{Vefff})
corresponds to a model with two massive scalar
(  2n-fermion  composite states)
bosons coupled to a massive vector
 (fermion-antifermion 
composite state)
 boson
built from the corresponding bilinears.
All the  scalar boson effective interactions
depend necessarily  on  the vector auxiliary field.

\section{Summary and conclusions}

Three effective  fermion models
were  investigated
by means of the auxiliary field method.
A minimal procedure was adopted to  introduce the minimum number 
of dynamical auxiliary fields and the minimum number of shifts to produce
the desired cancelation  of the fermion  interactions.
This reduces eventual ambiguities in the calculation.
In this minimal procedure it was assumed 
quite  strong coupling constants 
(except the one for the highest order coupling) 
with respect to  (normalized) auxiliary fields  that 
only  fluctuates weakly around the ground state,
therefore  being   weak
with respect to the condensates.
Possible extensions to lift the condition of weak fields were proposed,
being that they  yield the same final effective boson  model and factorization result.
The solution of the (coupled) gap equations corresponds to the solution of the 
first gap equation with however a strong dependence on the coupled expressions 
for the functions $\beta_n$, as presented in the case of the first model 
with expressions (\ref{betas}).
For larger number of fermion components ($N_r$)
 solutions of the gap equations
only can be found in higher dimensions. 
For the cases in which the gap equations present solutions several conclusions could be drawn.
For both models it was found that all the higher order operators and condensates factorize
 into the lowest order, 
i.e. $<(\bpsi_a \psi_a)^n> =  <\bpsi_a \psi_a>^n$.
One  exception was found for  the case 
a  constant shift  in  the lowest order auxiliary variable 
that was  considered
for the most general series (second one),
shift $A_1$, expression (\ref{shifts-n-1}),
 for $\xi_1 \to \xi_1 - B_1 \bpsi_a \psi_a - A_1$.
In this case the higher order condensates do not factorize into the lowest order one.
The shift $A_1$ corresponds to a non trivial overall subtraction 
 of the corresponding lowest order
condensate $\xi_1^{(0)}$.

As a second step, the fermion determinants of the models 
were expanded in powers of the (weak) auxiliary fields.
These resulting effective  models  describe composite fermion  states
and interactions,  being  therefore related to 
a  previous fermion dynamical model.
The resulting polynomial interaction terms 
 were found to have meaninful  different structures from  each other.
By comparing the resulting boson effective  models given by  expressions  (\ref{eff-I}) 
(or the limit presented in  expression (\ref{Vlargecondens})) and (\ref{masses-II})
it is seen that the former has a more symmetric shape.
Furthermore,  the field $\chi_1$ 
(and, analogously, the field $\xi_1$) can have  a different contribution for the overal model
from the contribution of the other fields $\chi_2$ and $\chi_3$, and $\chi_n$ in general, 
being   still more apparent for the case of the second and third models
for   $\xi_n$ and $\varphi_n$ respectively.
Therefore
it  might happen  that only a sector of the resulting boson effective model
presents a more specific symmetry instead of the full model.
This is the case of the third case analysed in Section \ref{sec:vector}.
The limit of progressively large condensates 
(progressively weak $\beta_n$) 
for the model of 
Section \ref{sec:esp} was shown to provide a quite simple
effective potential in expression (\ref{Vlargecondens}).
It was found to be invariant under  continuous
transformations that preserve the length $\sum_n (\chi_n)$,
and  also  discrete permutation transformations.
This invariance does not come out in the second model for more general series.
However in the last case, for the interactions 
of the form $\sum_j^N   (\bar{\psi}_a \gamma_\mu \psi_a)^{2j}$,
without the formation of vector
(or any other) condensates a similar symmetry 
appeared for the higher order auxiliary fields representing 2 and 4 particle
states $\varphi_2, \varphi_4$.
This last case represents a  local and 
momentum independent limit of a
effective potential of a 
theory of fermions interacting by means of vector field exchange
with the basic structure of Quantum Electrodynamics in that limit.
Therefore these results suggest  that different  higher order powers 
 of fermion bilinears
might yield  boson models for  N-fermion states
  with    (approximated) symmetry  depending on 
the terms considered in such series and
on the values of the original  coupling constants of the  fermion  model.
As discussed in the Introduction, 
although a renormalization group flow for the first series investigated in 
Section (\ref{sec:esp}) can yield a full series of the type of Section (\ref{sec:geral}), 
it is possible to figure out that the terms of the first series keep the approximate
resulting symmetry  while the other terms of the more general series
tend to break it. This scenario might be realistic depending on the resulting 
relative strength of the coupling constants of the more general series as 
commented above.
The cases of fermions 
with the corresponding symmetries for the   internal quantum numbers
 (such as SU(2) or SU(3)  flavor)
were outside the scope of this work.
Since the appearance of the approximate symmetry for 2n-fermion states 
in the bosonized model was obtained without considering 
a chiral symmetry in the departing model of Section (\ref{sec:esp})
it is concluded that the approximate degeneracy between these multifermion
scalar states is not due to a chiral symmetry.
In hadron physics, the lightest  scalar mesons with similar masses around 1 GeV 
 have seemingly different structures such as quark-antiquark and tetraquark content
\cite{scalars1,scalars2} with similar masses. This approximate degeneracy may 
correspond to the approximate symmetry of expressions
(\ref{eff-I})  and (\ref{Vlargecondens}). 
One might expect however  that for the phenomenological coupling constants of
the light scalar mesons (comparable to the effective model of Section
(\ref{sec:esp})) chiral symmetry should be a relevant symmetry to be considered.
Although the fermion-antifermion  channels might be
different they must be related to two fermions and tetrafermion states and interactions
by crossing or Fierz transformation. 
Furthermore it is interesting to note that the auxiliary boson fields defined above
correspond to a set of states.
In (non relativistic) cold atoms there are also 
different n-particle states (2, 3 or 4 fermion or boson states)
 that have been observed to 
have similar energies  (binding energies) 
\cite{cold-atoms1,cold-atoms2}. 
Although this is a 
non relativistic  system the resulting
symmetry corresponds to the 
one found in the present work, i.e. a degeneracy between states 
with different number of particles.
Several questions arise such as:
which kinds
 of resulting approximate symmetries is it possible to obtain
in the bosonized model by switching on and off
particular terms of the 
original higher order effective potential for 
a particular structure of fermion bilinears $\bpsi \Gamma \psi$ 
(where 
$\Gamma$ is one particular operator acting on spinor or other internal 
space)?

 \section*{Acknowledgements}

The author thanks  short discussions with F.S.
 Navarra, P. Bedaque and J. Helayel Neto,
and partial financial support by CNPq- Brazil 
and FAPEG-Goias, Brazil.



\begin{thebibliography}{9}





\bibitem{qft} 
C. Itzykson, J.B. Zuber, {\it Quantum Field Theory},
McGraw Hill (1985).


 \bibitem{osipov-hiller1}
A. A. Osipov and B. Hiller,
Eur. Phys. J. {\bf C 35},   223 (2004).

 \bibitem{osipov-hiller2}
A.A. Osipov, B. Hiller, J. Moreira and A.H. Blin,
Eur. Phys. J. {\bf C  46}, 225 (2006).


\bibitem{osipov-hiller-blin} A.A. Osipov, B. Hiller, and A.H. Blin,
Eur. Phys. J. {\bf A 49}, 14 (2013).


\bibitem{simmons1}
C. M. Bender, K. A. Milton, M. Moshe, S. S. Pinsky, L. M. Simmons Jr., Phys. Rev.
{\bf D 37}, 1472 (1988).

\bibitem{simmons2}
S. S. Pinsky, L. M. Simmons, Jr.,
Phys. Rev. {\bf D 38}, 2518 (1988).

\bibitem{chalmers}
G. Chalmers,
JHEP {\bf 03}, 001 (1998).


\bibitem{nitta-susy}
 M. Nitta, Nucl. Phys. {\bf B 711} 133 (2005).

\bibitem{darkmatter}
M. B. Krauss, S. Morisi, W. Porod and W. Winter, JHEP 02, 056 (2014).
 

\bibitem{cahill} K. Cahill,
Phys. Rev.{\bf  D 88}, 125014  (2013);
Erratum Phys. Rev. D 89, 029905 (2014)


\bibitem{6th-order-cm}
A.A. Katanin, 
J. Phys. {\bf A 46}, 045002 (2013).

\bibitem{braaten-hammer}
E. Braaten, H.-W. Hammer, Phys. Rept. {\bf 428}, 259 (2006).

\bibitem{bedaque-kolck}
P.F. Bedaque, U. van Kolck,
Annu. Rev. Nucl. Part. Sci. {\bf 52}, 339 (2002).



\bibitem{weinberg79} 
S. Weinberg, Physica {\bf A 96}, 327 (1979).

\bibitem{ChPT1} 
J. Gasser and H. Leutwyler, Ann. Phys. (N.Y.) {\bf 158}, 142
(1984).

\bibitem{ChPT2} 
S. Scherer, Prog. in Part. and Nucl. Phys. {\bf 64}, 1 (2010).

\bibitem{kleinert} H. Kleinert, 
Lectures presented at the Erice Summer Institute 1976, in 
Understanding the Fundamental Constituents of Matter,
Plenum Press, New York, ed. by A. Zichichi, 289 (1978).


\bibitem{e-w-EFT}
For example in:
G. Isidori, PoS {\it CD09}, 073  (2009).




\bibitem{platter-etal}
L. Platter, H.-W. Hammer, and U.-G. Meissner, 
Phys. Rev. {\bf A 70}, 052101 (2004).

\bibitem{factorization} See for example in:
R.~Thomas, T.~Hilger and B.~Kampfer,
Nucl.\ Phys.\ A {\bf 795}, 19 (2007). 
 
\bibitem{zong-etal} 
H. -s. Zong, D. -k. He, F. -y. Hou and W. -M. Sun, Int. J. Mod. Phys. {\bf A 23}, 1507 (2008).

\bibitem{nonfactorization}
A.~Gomez Nicola, J.~R.~Pelaez and J.~Ruiz de Elvira,
Phys. Rev.{\bf D 82}, 074012 (2010).

\bibitem{braghin-navarra}
F.L. Braghin, F.S. Navarra,   Phys. Rev. {\bf D  91}, 074008 (2015).

\bibitem{cold-atoms1}
For example in:
J. H. Gurian, P. Cheinet, P. Huillery, A. Fioretti, J. Zhao, P. L. Gould, D. Comparat, and P. Pillet
Phys. Rev. Lett. {\bf 108}, 023005 (2012).

\bibitem{cold-atoms2}
For example in:
F. Ferlaino, S Knoop, M. Berninger, W. Harm, J. P. D'Incao, H.-C. Nagerl, R. Grimm 
Phys. Rev. Lett. {\bf 102}, 140401 (2009).

\bibitem{scalars1}
For example in:
K.A. Olive et al. (Particle Data Group), Chin. Phys. {\bf C 38}, 090001 (2014).

\bibitem{scalars2}
G. 't Hooft, G. Isidori, L. Maiani, A.D. Polosa, V. Riquer,
Phys.Lett.B {\bf 662}, 424 (2008).

\bibitem{kashiwa}
T. Kashiwa, Phys. Rev. {\bf D 59}, 085002 (1999).



\bibitem{kleinert2011}
H. Kleinert, Electronic Journal of Theoretical Physics {\bf 8}, 57
(2011).



\bibitem{GN} 
D.J. Gross, A. Neveu, Phys. Rev. {\bf D 10}, 3235 (1974).

\bibitem{klevansky}
S.P. Klevansky, Rev. of Mod. Phys. {\bf 64}, 649 (1992).

\bibitem{berdnorz} 
Adam Bednorz,
Eur. Phys. J. {\bf C 73} 2654 (2013).


\bibitem{lp4}  
E. Baum, Nucl. Phys. {\bf B 266}, 547 (1986).


\bibitem{condmat} 
J.W. Negele, H. Orland, Quantum Many-Particle Systems, 
Addison-Wesley, 1988.




\bibitem{alkofer-reinhardt}
H. Reinhardt, R. Alkofer, Phys. Lett. {\bf B 207}, 482 (1988).

\bibitem{mosel} U. Mosel,
{\it Path Integrals in Field Theory: An Introduction},
Springer-Verlag Berlin Heidelberg, (2004).






\end{thebibliography}
\end{document}